\documentclass[reprint,groupedaddress,showpacs,preprintnumbers,nofootinbib,nobibnotes,amsmath,amssymb,aps,prd,floatfix,superscriptaddress]{revtex4-1}

\usepackage[colorlinks=false]{hyperref}
\usepackage{dcolumn}
\usepackage{bm}
\usepackage[normalem]{ulem}  

\usepackage{graphicx,color,xcolor}
\usepackage{enumitem}
\usepackage[utf8]{inputenc}
\usepackage{mathrsfs}

\def\mnras{Mon. Not. of the Royal Astron. Soc.} 	

\begin{document}

\title{Magnetic field distribution in magnetars}

\author{Debarati Chatterjee}\email{dchatterjee@lpccaen.in2p3.fr}
\affiliation{LPC, UMR6534, ENSICAEN, 14050 Caen, France}%
\author{J\'er\^ome Novak}\email{jerome.novak@obspm.fr}
\author{Micaela Oertel}\email{micaela.oertel@obspm.fr}
\affiliation{LUTH, Observatoire de Paris, PSL Research University,
  CNRS, Universit\'e Paris Diderot, Sorbonne Paris Cit\'e, 5 place
  Jules Janssen, 92195 Meudon, France}%

\date{\today}

\begin{abstract}
  Using an axisymmetric numerical code, we perform an extensive study
  of the magnetic field configurations in non-rotating neutron stars,
  varying the mass, magnetic field strength and the equation of
  state. We find that the monopolar (spherically symmetric) part of
  the norm of the magnetic field can be described by a single profile,
  that we fit by a simple eighth-order polynomial, as a function of
  the star's radius. This new generic profile applies remarkably well
  to all magnetized neutron star configurations built on hadronic
  equations of state. We then apply this profile to build magnetized
  neutron stars in spherical symmetry, using a modified
  Tolman-Oppenheimer-Volkov (TOV) system of equations. This new
  formalism produces slightly better results in terms of mass-radius
  diagrams than previous attempts to add magnetic terms to these
  equations. However, we show that such approaches are less accurate
  than usual, non-magnetized TOV models, and that consistent models
  must depart from spherical symmetry. Thus, our ``universal''
  magnetic field profile is intended to serve as a tool for nuclear
  physicists to obtain estimates of magnetic field inside neutron
  stars, as a function of radial depth, in order to deduce its
  influence on composition and related properties. It possesses the
  advantage of being based on magnetic field distributions from
  realistic self-consistent computations, which are solutions of
  Maxwell's equations.
\end{abstract}


\maketitle

\section{Introduction}\label{s:intro}

The macroscopic structure and observable astrophysical properties of
neutron stars depend crucially on its internal composition and thus
the properties of dense matter. The Equation of State (EoS) determines
global quantities such as observed mass and radius. Transport 
properties such as thermal conductivity and bulk viscosity have an 
effect on cooling observations as well as emission of gravitational
waves. As we enter an era of multi-messenger astronomy, it is crucial 
to construct consistent microscopic and macroscopic models in order to 
correctly interpret astrophysical observations.

There are a large number of astrophysical observations, \textit{e.g.}
soft-gamma repeaters (SGR) or anomalous X-ray pulsars (AXP), that
indicate the existence of ultra-magnetized neutron stars or
magnetars~\cite{kaspi-17}. While such observations only probe the
surface magnetic field, there is no way to measure directly the
maximum magnetic field in the interior. Using the simple virial
theorem, one may estimate the maximum interior magnetic field to be as
high as $10^{18}$~G. If such large fields exist in the interior, they
may strongly affect the energy of the charged particles by confining
their motion to quantized Landau levels and consequently modify the
particle population, transport properties as well as the global
structure \cite{avancini-18, tolos-17, franzon-16b, franzon-16,
  gomes-18, gomes-17b, gomes-17, gomes-14, gomes-13, dexheimer-12,
  dexheimer-14, wei-17b, wei-17}. However, it is necessary to know the
magnetic field amplitude at a given location in the star,
\textit{i.e.} a magnetic field distribution, in order to determine its
effect on the internal composition and EoS.

The ideal way to tackle that problem would of course be to
self-consistently solve the neutron star structure equations endowed
with a magnetic field, \textit{i.e.} combined Einstein, Maxwell and
equilibrium equations, together with a magnetic field dependent EoS,
as done by Chatterjee et al.~\cite{chatterjee-15}. This solution is
complicated by the fact that in presence of a magnetic field, the
neutron star structure strongly deviates from spherical symmetry and
the spherically symmetric Tolman-Oppenheimer-Volkov (TOV) equations
are no longer applicable for obtaining the macroscopic structure of a
the neutron star \cite{bocquet-95,chatterjee-15, dexheimer-17c}. For
small magnetic fields, perturbative solutions have been
developed~\cite{konno-99}, but can no longer be applied for field
strengths which might influence matter properties.

There have been several attempts to determine neutron star structure
assuming an \textit{ad hoc} profile of the magnetic field, without
solving Maxwell's equations, within the TOV system (see
\textit{e.g.}~\cite{casali-14, sotani-17, chu-18}). To that end, many
authors employ the parameterization introduced twenty years ago by
Bandyopadhyay et al. \cite{bandyopadhyay-97}, where the variation of
the magnetic field norm $B$ with baryon number density $n_B$ from the
centre $B_c$ to the surface $B_s$ of the star is given by the form
\begin{equation}
B (n_B/n_0) = B_s + B_c [1 - \exp( - \beta (n_B/n_0)^{\gamma})]~,
\label{eq:bprofile1}
\end{equation}
with two parameters $\beta$ and $\gamma$, chosen to obtain the desired
values of the maximum field at the centre and at the surface. This is
an arbitrary profile, which possesses the same symmetries as the
baryon density distribution in the star. Parameters $\left( \beta,
  \gamma \right)$ are chosen such that the surface field is consistent
with observations and the maximum field prevailing at the center
conforms to the virial theorem. 

Lopes and Menezes \cite{lopes-15} later introduced a variable
magnetic field, which depends on the energy density rather than on the
baryon number density:
\begin{equation}
B = B_c \left( \frac{\epsilon_M}{\epsilon_0} \right)^{\gamma} + B_s~,
\label{eq:bprofile2}
\end{equation}
where $\epsilon_M$ is the energy-density of the matter alone,
$\epsilon_0$ is the central energy density of the maximum mass
non-magnetic neutron star and a parameter $\gamma> 0$, arguing that
this formalism reduces the number of free parameters from two to
one. The authors put forward as additional motivation the fact that it
is the energy density and not the number density that is relevant in
TOV equations for structure calculations. To account for anisotropy in
the shear stress tensor, they applied the above field profile in a
formalism~\cite{bednarek-03}, where the different elements containing
the pressure are ``averaged'', leading to shear stress tensor of the
form diag($B^2/24\pi, B^2/24\pi, B^2/24\pi$) \cite{dexheimer-12,
  menezes-16}. Nevertheless, this approach within the TOV system is
still spherically symmetric and the parameter $\gamma$ not related to
any experimental or observational constraint. There have also been
suggestions of the magnetic field profile being a function of the
baryon chemical potential \cite{dexheimer-12} as:
\begin{equation}
B(\mu_B) = B_s + B_c \left[ 1-\exp(b \frac{(\mu_B-938)^a}{938} ) \right]~,
\end{equation}
with $a=2.5$, $b=-4.08 \times 10^{-4}$ and $\mu_B$ given in MeV. In
contrast to the profiles in
Eqs.~(\ref{eq:bprofile1},\ref{eq:bprofile2}), such a formula avoids
that a phase transition induces a discontinuity in the effective
magnetic field. 

However, it was subsequently pointed out by Menezes and Alloy
\cite{menezes-16b} that any of the above \textit{ad hoc} formulations
for magnetic field profiles are physically incorrect since they do not
satisfy Maxwell's equations. In particular, it is obvious that
assuming such a magnetic field profile in a spherically symmetric star
implies a purely monopolar magnetic vector field distribution, which
is incorrect. The inconsistency of this type of approach can be seen,
too, by inspecting the most general solution of the equations of
hydrostatic equilibrium in general relativity for a spherically
symmetric star. In Schwarzschild coordinates,
$\left(\bar{t}, \bar{r}, \bar{\theta}, \bar{\varphi} \right)$, the
line element reads: 
\begin{equation}
  \label{e:def_Schwarzschild}
  {\rm d}s^2 = -e^{-2\Phi}\, {\rm d}\bar{t}^2 + \left( 1 -
    \frac{2Gm}{\bar{r}} \right)^{-1}\hspace{-1em} {\rm d}\bar{r}^2
  + \bar{r}^2 \left( {\rm d}\bar{\theta}^2 + \sin^2\bar{\theta}
    {\rm d}\bar{\varphi}^2 \right),
\end{equation} 
where $m(\bar r)$ and $\Phi(\bar r)$ are the two relativistic
gravitational potentials defining the metric (at the Newtonian limit,
$m$ represents the total mass enclosed in the sphere of radius
$\bar r$, and $\Phi / c^2$ becomes the Newtonian gravitational
potential). The resulting coupled system of equations for the star's
structure has been derived by Bowers and Liang~\cite{bowers-74} and
reads
\begin{eqnarray}
  \frac{dm}{d\bar r} &=& 4\pi \bar r^2 \varepsilon \nonumber\\
  \frac{d\Phi}{d\bar r} &=& \left( 1 - \frac{2Gm}{\bar rc^2} \right)^{-1} \left(
                            \frac{Gm}{\bar r^2} + 4\pi G\frac{p_r}{c^2}\bar
                            r \right) \nonumber\\ 
  \frac{dp_r}{d\bar r} &=& -\left(\varepsilon +
                           \frac{p_r}{c^2} \right) \frac{d\Phi}{d\bar r} +
                           \frac{2}{\bar r} (p_\perp - p_r) ~, 
                           \label{eq:bowers}
\end{eqnarray} 
with an energy-momentum tensor of the form
$T^{\mu\nu} = \mathrm{diag}(\varepsilon, p_r, p_\perp,p_\perp)$, where
$p_r$ and $p_{\perp}$ are the radial and tangential pressure
components. This is the most general energy-momentum tensor one can
use assuming spherical symmetry and it goes beyond the perfect-fluid
model, for which $p_r = p_\perp$. One may be tempted to cast a
general electromagnetic energy-momentum tensor assuming a perfect
conductor and isotropic matter, and for a magnetic field pointing in
$z$-direction (see \textit{e.g.}~\cite{chatterjee-15}) into this
form. However, in the case of the electromagnetic energy-momentum
tensor $T^{\theta\theta} \neq T^{\phi\phi}$ (look at Eqs.~(23d)-(23e)
of~\cite{chatterjee-15}), in clear contradiction with the assumption
of Bowers and Liang~(\ref{eq:bowers}) in spherical symmetry. Another
problem arises from the fact that
$\lim_{r\to 0} (T^{rr} - T^{\theta\theta}) \not= 0$ and thus, the last
term in Eq.~(\ref{eq:bowers}) diverges at the origin (from the first
line in this equation, one sees that the quantity $m(\bar r)\sim \bar r^3$ and
therefore $m/\bar r^2$ does not diverge). This discussion
shows that there cannot be any correct description of the magnetic
field in spherical symmetry.

Starting from two-dimensional numerical models, Dexheimer \textit{et
  al.}~\cite{dexheimer-17b,dexheimer-17} performed a fit to the shapes of the
magnetic field profiles following the stellar polar direction as a
function of the chemical potentials (as in \cite{dexheimer-12}) by
quadratic polynomials instead of exponential ones as
\begin{equation}
B(\mu_B) = \frac{(a+b\mu_B + c\mu_B^2)}{B_c^2} \mu ~,
\end{equation}
where $a,b,c$ are coefficients determined from the numerical
fit. Unfortunately, no check of the validity of this fit has been
shown in these works for other directions. In Ref.~\cite{mallick}, a
density dependent profile is applied within a perturbative
axisymmetric approach à la Hartle and Thorne~\cite{hartle-68}, but
without solving Maxwell's equations. It remains, however, that the
star's deformation due to the magnetic field implies that such a
density (or equivalent) dependent profile depends on the direction,
thus will be different looking \textit{e.g.}  in the polar or the
equatorial direction.

In view of all these intrinsic difficulties, we will not propose here
a simple scheme for solving structure equations of magnetized stars --
to that end we refer to the publicly available numerical codes
assuming {\em axial\/} symmetry~\cite{chatterjee-15,xns}. Instead,
since in many cases it might be sufficient to have an idea of the
order of the value of the magnetic field strength to test its
potential effect on matter properties, our aim is to provide a
``universal'' magnetic field strength profile from the surface to the
interior obtained from the field distribution in a fully
self-consistent numerical calculation from one of these
codes. Further, we probe the applicability of this profile for
determining the structure of magnetized neutron stars in an
approximate way in spherical symmetry compared with full numerical
structure calculations. As we will show, qualitatively the correct
tendency can be reproduced for some NS properties, but to reproduce
quantitatively correct results, the full solution has to be applied.

The paper is organized as follows. Sec.~\ref{s:formalism} describes
our physical models, including the EoSs we use in this manuscript,
together with the numerical techniques applied to solve the
models. Sec.~\ref{s:results} provides the magnetic field profiles
derived numerically by varying certain physical parameters, to achieve
a generic profile for the monopolar part of the norm of the magnetic
field. This profile is then applied in Sec.~\ref{s:TOV} to a modified
TOV system, to see its effect on NS masses and radii. Finally,
Sec.~\ref{s:conc} gives a summary of our work, together with some
concluding remarks.

\section{Formalism and models}\label{s:formalism}

In this section, we summarize the numerical approach for
self-consistently modelling magnetized neutron stars. More details
can be found in in~\cite{chatterjee-15,bocquet-95,chatterjee-17}.

\begin{table*}[t]
\begin{center}
\begin{tabular}{l|cccccccc}
\hline
 Model& $n_{\rm sat}$  & $E_B$ & $K$ & $E_{\rm sym}$& $L$& $ M_\mathit{max}$ & $R_{1.4}$& $\tilde{\Lambda}(q = .8)$  \\
&$(\mathrm{fm}^{-3})$ & (MeV) & (MeV) &(MeV) & (MeV)&($M_\odot$) &(km) & \\ \hline \hline
HS(DD2)         &0.149   & 16.0 & 243& 31.7 &55&2.42 &13.2 &810 \\
SFHoY& 0.158&16.2 & 245&31.6&47& 1.99&11.9 &399 \\
STOS & 0.145&16.3 & 281&36.9&111&2.23 &14.5 &1420\\
BL\_EOS& 0.17&15.2 & 190&35.3&76&2.08 & 12.3&466 \\
SLy9& 0.15&15.8 & 230&32.0&55&2.16 &12.5 &533\\
SLy230a& 0.16&16.0 & 230&32.0&44& 2.11&11.8 &401\\
\hline

\end{tabular}
\caption{Saturation density , $n_{\rm sat}$, the binding energy at
  saturation $E_B$, the compression modulus $K$, the symmetry energy,
  $E_{\rm sym}$ and its slope, $L$, of symmetric nuclear matter are
  listed for the different EoS models employed. The neutron star
  maximum gravitational mass and radius at a fiducial mass of
  $M_G = 1.4 M_\odot$ are given for cold spherical
  stars. $\tilde{\Lambda}$ is the tidal deformability during inspiral
  of a binary neutron star merger, calculated with the chirp mass as
  measured for GW170817~\cite{ligo-17,abbott-18}. $\tilde{\Lambda}$ is
  only very weakly dependent on the mass ratio of the two stars,
  $q = M_1/M_2$ and is given here for a reference value of $q = 0.8$
  (the same EoS is assumed for both stars). Note that the STOS EoS is
  as well excluded by the constraints on $\tilde\Lambda$ obtained by
  GW170817~\cite{abbott-18} as by nuclear physics experiments
  indicating much lower values for the slope of the symmetry
  energy~\cite{oertel-17}. We will nevertheless keep this EoS model as
  representative of an extreme case with large neutron star radius and
  symmetry energy slope.}
\label{tab:nmatter}
\end{center}
\end{table*}

\subsection{Non-rotating magnetized neutron stars in general
  relativity}\label{ss:magNS}

Due to the high compactness of neutron stars, we consider models within
the theory of general relativity and solve coupled Einstein-Maxwell
partial differential equations. We follow the scheme described
in~\textcite{bonazzola-93}, who considered the general case of
rotating neutron stars, with the assumptions of stationarity, axial
and equatorial symmetry, and circular spacetime, where the metric is
given in the quasi-isotropic gauge, different from that used in TOV
systems~(\ref{e:def_Schwarzschild}), by:
\begin{eqnarray}
  {\rm d}s^2 &=& -N^2\, {\rm d}t^2 + C^2r^2\sin^2\theta \left({\rm d}\varphi -
    N^\varphi\, {\rm d}t \right)^2 \nonumber\\
  &&+ A^2\left( {\rm d}r^2 + r^2\, {\rm d}\theta^2\right),\label{e:def_metric}
\end{eqnarray}
where $N, N^\varphi, A$ and $C$ are the relativistic gravitational
potentials which are, as all other fields in
Secs.~\ref{s:formalism}-\ref{s:results}, functions of the coordinates
$(r,\theta)$ only (independent from the $\varphi$-coordinate).

In this work, we shall restrict ourselves to the case without
rotation, which in particular implies that there is no electric field
in the models (perfect conductor). Nevertheless, as said in the
introduction, the presence of a magnetic field induces a distortion of
the stellar structure, which cannot remain spherically symmetric. Due
to spacetime symmetries and circularity condition, only two magnetic
field geometries can be described within this framework: a purely
poloidal magnetic field (see~\textcite{bocquet-95}) or a purely
toroidal one (see \textcite{kiuchi-08} and later
\textcite{frieben-12}). In this work, we consider only purely poloidal
magnetic fields, meaning that the only non-trivial components are
$B^r(r, \theta)$ and $B^\theta(r, \theta)$. This choice results in an
asymptotically dipolar magnetic field distribution.

Matter is supposed to be composed of a perfect fluid coupled to the
magnetic field. In \textcite{chatterjee-15}, it has been shown that
the use of a magnetic field dependent EoS and inclusion of
magnetization in the equations have negligible effects on neutron star
structure, at least up to a polar magnetic field
$B_{\rm pole} \sim 5 \times 10^{17}$~G (roughly corresponding to to a
central magnetic field value $b_c \sim 2\times 10^{18}$~G) with a
simple quark model EoS. We therefore neglect magnetic field dependency
of the EoS and magnetization here, but they could be included in a
straightforward way~\cite{chatterjee-15}. Matter is also assumed to be
perfectly conducting and the magnetic field originates from free
currents, moving independently from the perfect fluid. Equilibrium
equations are obtained from the divergence-free condition of the
energy-momentum tensor, and can be written as a first integral of
motion, ~\cite{bonazzola-93, chatterjee-15}. It is mostly the Lorentz
force term in this equilibrium equation which distorts the stellar
structure and makes it deviate from spherical symmetry. To summarize,
given an equation of state (EoS) for nuclear matter (see
Sec.~\ref{ss:EOS} hereafter), we thus solve the system of coupled
Einstein-Maxwell equations, together with magnetostatic
equilibrium. These models are then characterized by their
gravitational mass ($M_G$, see~\cite{bonazzola-93} for a definition),
their EoS (see Sec.~\ref{ss:EOS}) and the central magnetic field,
$b_c$.
 
\subsection{Numerical methods}\label{ss:num}

The equations to be solved to get axisymmetric solutions form a set of six
non-linear elliptic (Poisson-like) partial differential equations,
coupled together with non-compact support (sources for
gravitational field extend up to spatial infinity). These equations
are solved using the same procedure as described in
\textcite{bocquet-95}, employing the numerical library
\textsc{lorene}~\cite{lorene} based on spectral methods for the
representation of fields and the resolution of partial differential
equations (see \textcite{grandclement-09}). 

Numerical
accuracy of the axisymmetric solutions is checked through an
independent test, the so-called \emph{relativistic virial theorem}
(\textcite{bonazzola-94, gourgoulhon-94}). This gives an upper bound
on the relative accuracy of the obtained numerical solution, and we
checked that it always remained lower than $10^{-4}$ for the
axisymmetric models presented in Sec.~\ref{s:results}.

\subsection{Equations of state}\label{ss:EOS}

The system of equations described above is closed by the EoS for
nuclear matter relating the pressure $p$ to the baryon density $n_B$.
Our selection of EoSs for the present work has been guided by the idea
to represent a large variety of different neutron star compositions
and nuclear properties, derived from completely different nuclear
physics formalisms. This was done in order to achieve an unbiased
universal parameterization applicable to any realistic nuclear EoS.
We consider one EoS model resulting from a microscopic calculation
(``BL\_EOS'') \cite{bombaci-18}\footnote{The model calculation exist
  only for homogeneous matter and a crust has been added, see the
  CompOSE entry for details.}. It uses the Brueckner-Hartree-Fock
formalism to tackle the many-body-problem and employs chiral
interactions for the basic two- and three-body nuclear
interactions. In addition, we consider several phenomenological mean
field models. These models are unified models in the sense that the
crust EoS has been obtained with the same nuclear interaction than the
one for the core guaranteeing consistency at the crust-core
transition. They include two non-relativistic Skyrme parameterizations
(``SLy9''and ``SLy230a'')\cite{chabanat-95,chabanat-97} with the crust
model from \textcite{gulminelli-15}, two relativistic mean field
models (``STOS'' and ``HS(DD2)'')~\cite{shen-98,typel-09} with the
crust obtained from the model in \textcite{hempel-09}, supposing a
temperature of 0.1 keV. The former one contains non-linear
interactions whereas the latter one is constructed with
density-dependent couplings. One model with hyperons (``SFhoY'')
\cite{fortin-17} completes our list of EoS models. It is a nonlinear
model where the crust is again obtained from
Ref.~\cite{hempel-09}. Hyperonic interactions have been chosen to
correctly reproduce hypernuclear data and a neutron star maximum mass
above current observational limits. Some nuclear and neutron star
properties of the different EoS models are listed in
Table~\ref{tab:nmatter}. All EoS data are available from the on-line
database CompOSE~\cite{compose}.

\begin{figure}[h]
  \center
  \includegraphics[angle=-90,width=0.8\columnwidth]{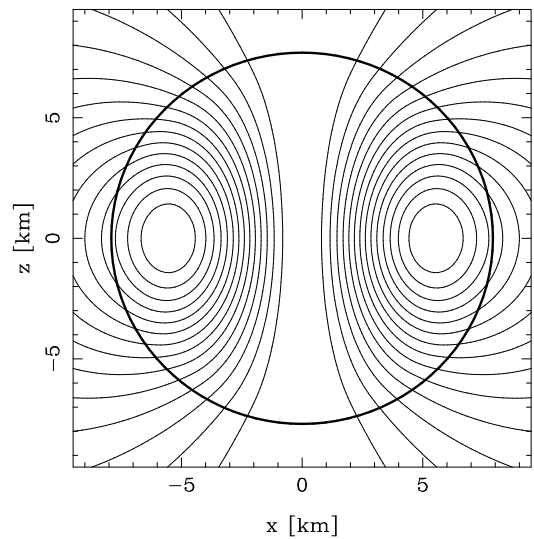}
  \caption{Magnetic field lines in the $(x,z)$-plane for a
    $M_G = 2\ M_\odot$ neutron star model endowed with a magnetic
    field which central value is $b_c = 5 \times 10^{17}\ {\rm G}$ and
    using the SLy203a EoS of Tab.~\ref{tab:nmatter}. Thick line
    denotes the surface of the star and the magnetic moment is along
    the $z$-axis.}
  \label{f:Bfield}
\end{figure}

\section{Generic magnetic field profile}\label{s:results}

The numerical models of neutron stars endowed with a magnetic field
described in Sec.~\ref{ss:magNS} consider two components ($B^r$ and
$B^\theta$) of the magnetic field vector, as measured by the Eulerian
observer (see \textcite{bocquet-95} for details). In the case of
non-rotating stars considered here, this magnetic field is the same as
that measured in the fluid rest-frame, denoted as $b^r$ and $b^\theta$
in \textcite{chatterjee-15}. As an example, the magnetic field
distribution of a full neutron star model is displayed in
Fig.~\ref{f:Bfield}, for a central value of the magnetic field
$b_c = 5\times 10^{17}$~G. The surface of the star (thick line) does
not exhibit any significant deviation from spherical shape, but it is
clear that the magnetic field distribution is dominated by the dipolar
structure and cannot be accurately described by any
spherically-symmetric model.

When trying to parameterize the magnetic field profile, the simplest
approach is to consider the norm of the magnetic field, namely
\begin{equation}\label{e:norm_b}
b = \sqrt{g_{rr}\left(b^r\right)^2 + g_{\theta\theta} \left( b^\theta
  \right)^2} = A\sqrt{\left(b^r\right)^2 + \left( b^\theta \right)^2}, 
\end{equation}
where the relativistic gravitational potential $A(r,\theta)$ has been
defined in Eq.~(\ref{e:def_metric}). Note that $b$ is the quantity
that enters the EoSs which take into account magnetization, as
explained \textit{e.g.} in~\cite{chatterjee-15}. The central value of
this magnetic field norm is denoted as $b_c = b(r=0)$ (independent of
$\theta$). In the rest of this work, we will consider this field as
the main object of our study.

\begin{figure}[h]
  \includegraphics[width=1.1\columnwidth]{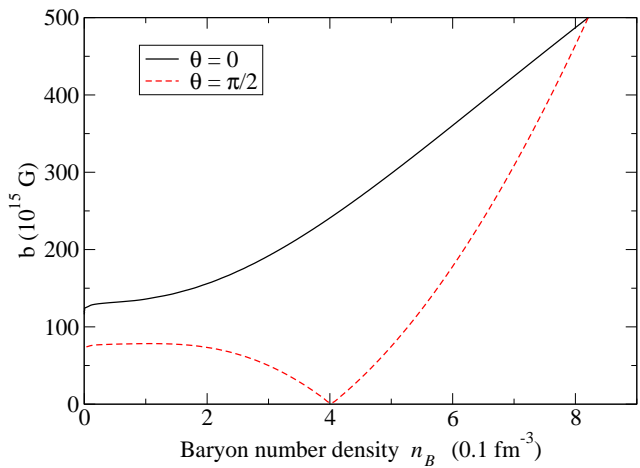}
  \caption{Magnetic field norm $b$~(\ref{e:norm_b}) as a function of
    the baryon density for two angular directions for the same stellar
    model as in Fig.~\ref{f:Bfield}.}
  \label{f:Bvsnb_frf}
\end{figure}

As stated in the introduction, several authors have considered a
parameterization of the magnetic field norm by the baryon density
$n_B$. In Fig.~\ref{f:Bvsnb_frf} we have plotted, for the same neutron
star model of $M_G=2\ M_\odot$ and $b_c = 5\times 10^{17}$~G as in
Fig.~\ref{f:Bfield}, the norm of the magnetic field as a function of
baryon number density $n_B$, along two radial directions: for
$\theta=0$ (passing through the pole) and for $\theta=\pi/2$ (passing
through the equator). As these two curves show noticeable differences,
including close to the center of the star
($n_B \sim 0.8\ {\rm fm}^{-3}$), it seems that this type of
parameterization can induce some inconsistency when describing
magnetic field in a neutron star. We therefore try to improve it and
adopt a different approach, taking a multipolar expansion of the
magnetic field norm ($Y_\ell^m(\theta, \varphi)$ being the spherical
harmonic functions):
\begin{equation}
  \label{e:b_ylm}
  b (r, \theta) \simeq \sum_{\ell =0}^{L_{\rm max}} b_\ell (r) \times
  Y_\ell^0(\theta).
\end{equation}
\begin{figure}[h]
  \includegraphics[height=1.1\columnwidth,angle=-90]{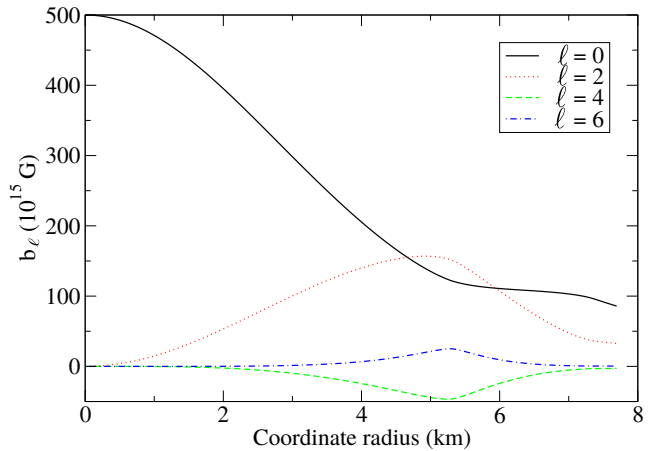}
  \caption{Radial profiles of the first four even multipoles,
    $b_\ell, \ell =0, 2, 4$ and $\ell = 6$, see definition
    (\ref{e:b_ylm}), of the magnetic field norm $b(r,\theta)$ computed
    for the stellar model described in Fig.~\ref{f:Bfield}. From
    symmetry arguments odd multipoles are all zero.}
  \label{f:multi_l}
\end{figure}

In Fig.~\ref{f:multi_l} we have plotted the first four non-zero terms
of this multipolar decomposition as functions of the coordinate radius
$r$. Note that, because of the symmetry with respect to the equatorial
plane, odd-$\ell$ terms in the decomposition (\ref{e:b_ylm}) are all
zero. It appears that, at least in the high-density central regions of
the star, the monopolar term $b_0(r)$, which is spherically symmetric,
is dominant over the others. It is important to stress here that,
contrary to the magnetic (vector) field, which has no monopolar
part in terms of vector spherical harmonics, the norm of the vector
field considered here is a scalar field which can possess a monopolar
component.

\begin{figure*}[t]
  \begin{center}
    \includegraphics[width=\columnwidth]{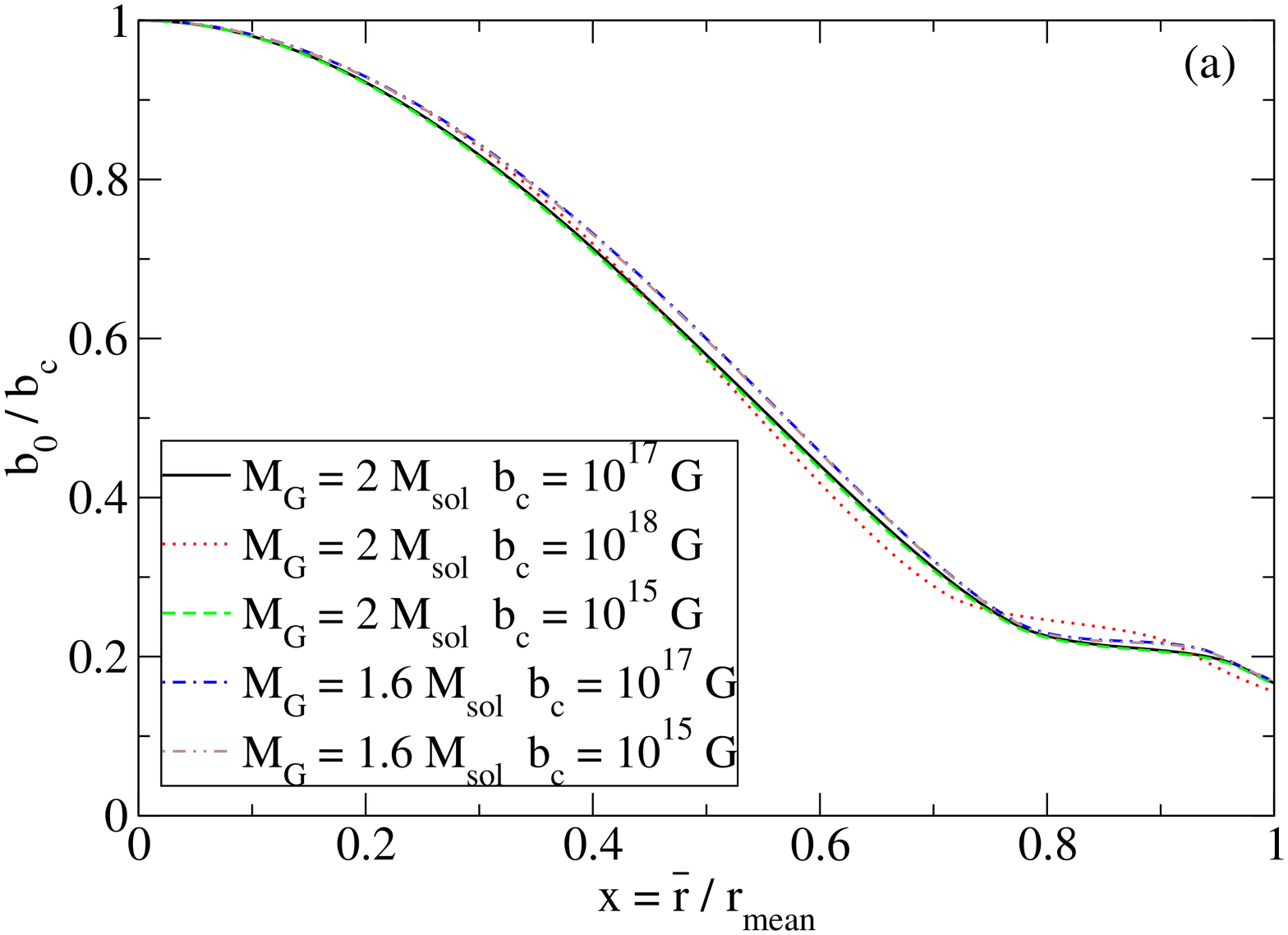}\hfill
    \includegraphics[width=\columnwidth]{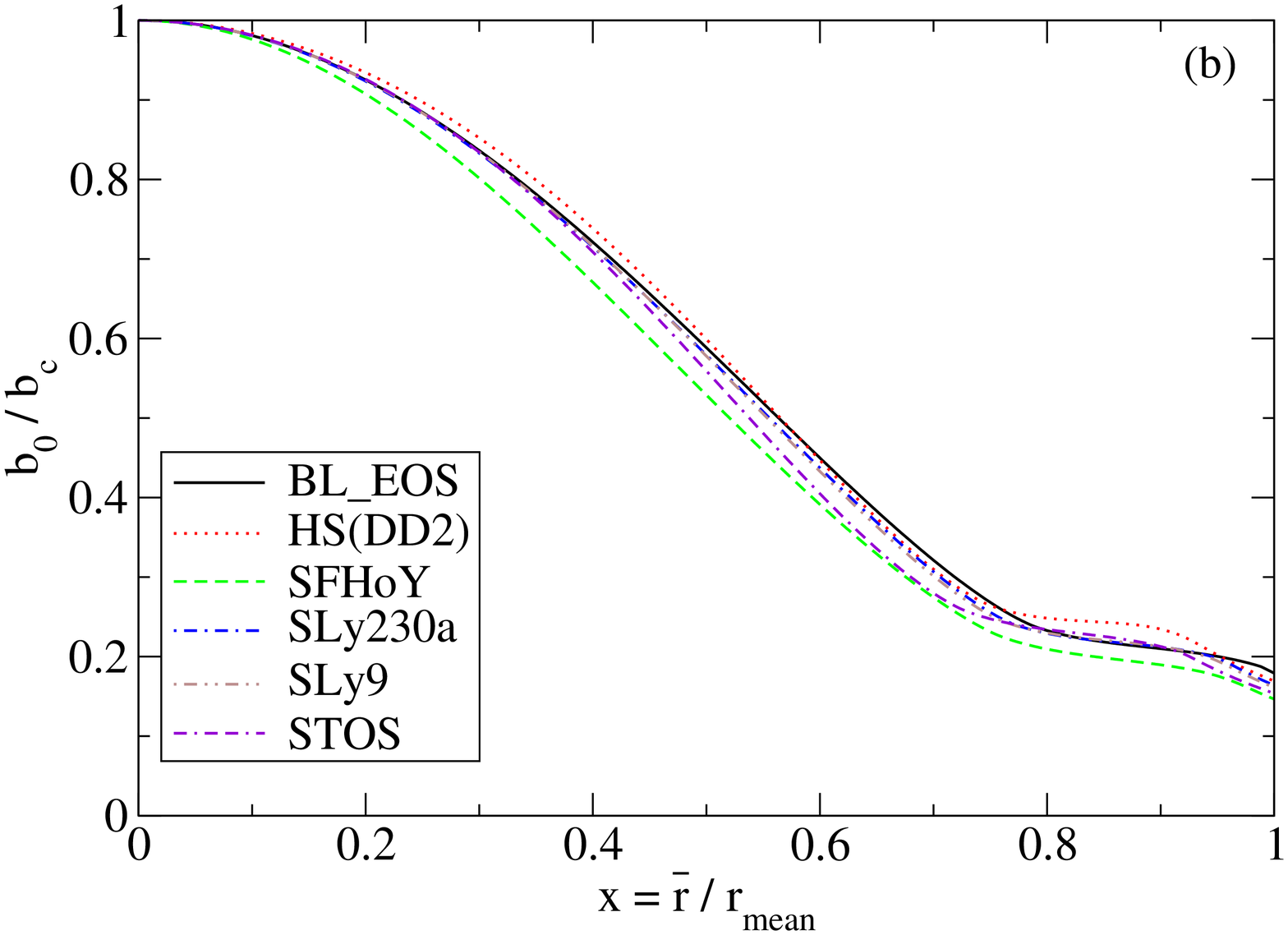}
    \caption{Monopolar part of the magnetic field profile $b_0(r)$,
      normalized to its central value for different magnetized
      neutron star models, as functions of the radius expressed in
      Schwarzschild coordinates $\bar{r}$ (see
      expression~(\ref{e:def_Schwarzschild})) divided by the star's
      mean radius (see text). \textit{Left panel (a):} all models are
      using SLy230a EoS (see Tab.~\ref{tab:nmatter}) but have
      different masses ($1.6\ M_\odot$ or $2\ M_\odot$) and
      different central magnetic fields $b_c$
      ($10^{15}\ \rm{G}, 10^{17}$~G or $10^{18}$~G). \textit{Right
        panel (b):} all are $2\ M_\odot$ models, with a central magnetic
      field $b_c = 5 \times 10^{17}$~G, but with different EoSs, see
      Tab.~\ref{tab:nmatter} for details.}
    \label{f:uprof}
  \end{center}
\end{figure*}

We then look at the behavior of the radial profile of $b_0$ when
varying the neutron star model in Fig.~\ref{f:uprof}. On the left
panel, we vary the gravitational mass of the star (either
$1.6\ M_\odot$ or $2\ M_\odot$), as well as the amplitude of the
magnetic field central value $b_c$ ($10^{15}$~G, $10^{17}$~G and
$10^{18}$~G). On the right panel of Fig.~\ref{f:uprof}, we vary the
EoS used in the stellar model, keeping the gravitational mass to be
$2\ M_\odot$ and the central magnetic field $b_c = 5\times
10^{17}$~G. These profiles are no longer displayed as functions of the
quasi-isotropic coordinate radius $r$, defined by the line
element~(\ref{e:def_metric}), but in view of the application to
TOV-systems in Sec.~\ref{s:TOV}, we consider here the Schwarzschild
coordinate radius $\bar{r}$, defined by the line
element~(\ref{e:def_Schwarzschild}). The gauge transformation is
obtained numerically and profiles are displayed as functions of this
radius divided by the star's mean radius $r_{\rm mean}$ which is such
that the integrated (coordinate-independent) surface of the star reads
$\mathcal{A} = 4\pi r_{\rm mean} ^2$. Indeed, when the star gets
distorted because of the magnetic field, it is difficult to define
uniquely a relevant radius. In that sense, $r_{\rm mean}$ is directly
connected to the star's surface and some of its emission properties.

It is remarkable that, although all possible parameters defining a
magnetized stellar model (mass, central magnetic field, EoS) have
been varied, all profiles are quite similar and deviate one from
another only by a few percent. The only case where a noticeable
difference appears is when using quark matter EoS. Therefore, we make
the following conjecture: the monopolar part of the norm of the
magnetic field follows a universal profile, up to minor variations,
when considering different neutron star models with realistic hadronic
EoSs. This ``universal'' profile has been fitted using a simple
polynomial:
\begin{equation}
  \label{eq:bprofile}
  b_0(x) = b_c \times \left(1 - 1.6x^2 -x^4 + 4.2x^6 -2.4x^8 \right),
\end{equation}
where $x = \bar{r} / r_{\rm mean}$ is the ratio between the radius
$\bar{r}$ in Schwarzschild coordinates~(\ref{e:def_Schwarzschild}) and
the star's mean (or areal) radius. Let us stress that the aim of the
present investigation is to obtain a universal profile for realistic
EoSs and that we have therefore excluded polytropic EoSs. A preliminary
calculation showed that the general parameterization applicable to the
family of realistic EoSs is not applicable directly to the case of
polytropes without specific fine tuning.

\begin{figure*}
  \begin{center}
      \includegraphics[width=\columnwidth]{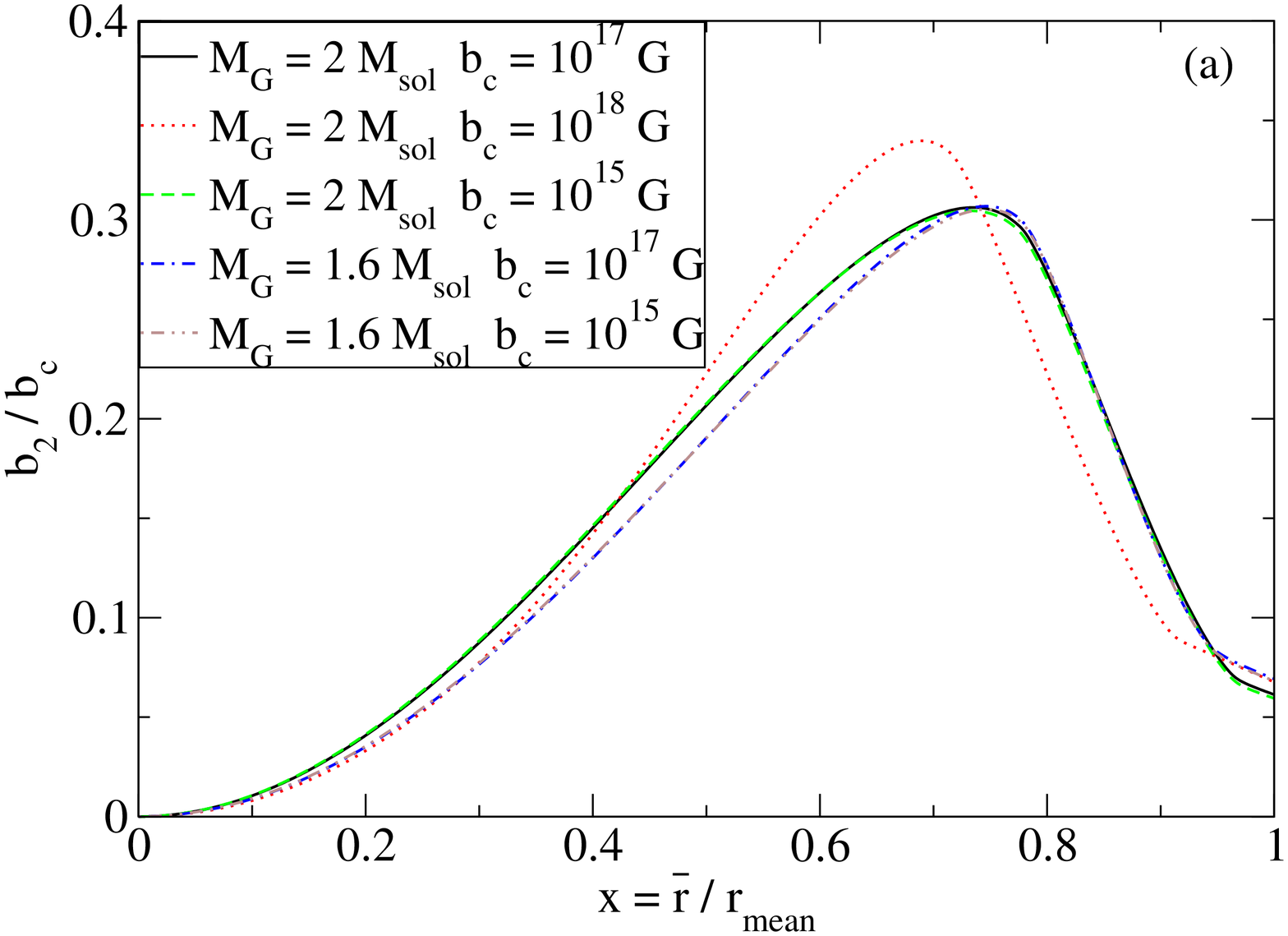}\hfill
      \includegraphics[width=\columnwidth]{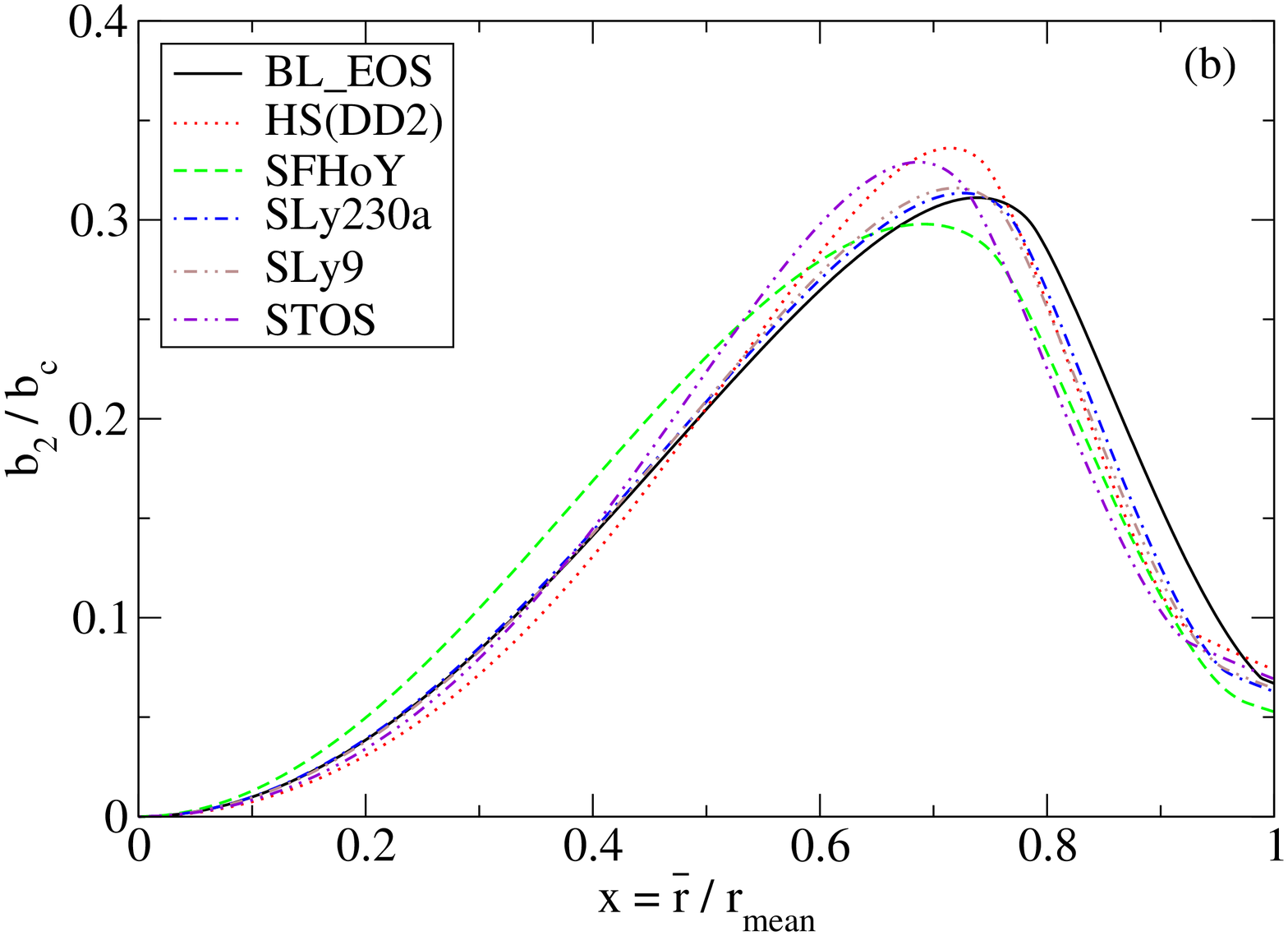}
      \caption{Same as Fig.~\ref{f:uprof} but for the dipole term
        $b_2(r)$ of the multipolar expansion~(\ref{e:b_ylm}) of the
        magnetic field norm.}
    \label{f:uprofl2}
  \end{center}
\end{figure*}

Going further, we display in Fig.~\ref{f:uprofl2}{} profiles for the
dipolar part $b_2(r)$, defined in Eq.~(\ref{e:b_ylm}). A larger
dispersion is visible in these plots, in particular in terms of EoSs
(right panel of Fig.~\ref{f:uprofl2}) and for the largest value of
magnetic field ($b_c = 10^{18}$~G, left panel of
Fig.~\ref{f:uprofl2}). This last point can be understood from the
large deformation undergone by the star at this value of central
magnetic field, when the contribution from higher-order multipoles
starts to become important. Together with Fig.~\ref{f:multi_l}, these
curves show that spherical symmetry is not a good approximation for
strongly magnetized neutron stars, where higher multipoles can have a
non-negligible effect. However, the relative robustness of the
profiles shown in Fig.~\ref{f:uprofl2} indicates that, for not too
large magnetic fields, the dipolar correction may be included in an
EoS-independent way and some perturbative approach to spherical
symmetry may be devised, in a way similar to \textcite{konno-99}. We
leave these investigations to further studies.

\section{Application to a TOV-like system}\label{s:TOV}

As discussed in the introduction, it is fundamentally inconsistent to
solve spherically symmetric equations for magnetized neutron star
models since it completely neglects the star's deformation due to the
electromagnetic field. It is, however, tempting, to have a simple
approach at hand which allows at least to qualitatively reproduce the
effects of the magnetic field on (some) neutron star properties
performing calculations only slightly more complicated than solving
TOV equations. To that end, we modify the TOV system by adding the
contribution from the magnetic field to the energy density and a
Lorentz force term to the equilibrium equation:
\begin{eqnarray}
  \frac{dm}{d\bar r} &=& 4\pi \bar r^2 (\varepsilon +
                         \frac{b^2}{\mu_0}) \nonumber\\ 
  \frac{d\Phi}{d\bar r} &=& \left( 1 - \frac{2Gm}{\bar rc^2} \right)^{-1} \left(
                            \frac{Gm}{\bar r^2} + 4\pi
                            G\frac{p}{c^2}\bar r \right) \nonumber\\ 
  \frac{dp}{d\bar r} &=& -\left(\varepsilon + \frac{b^2}{\mu_0} +
                           \frac{p}{c^2} \right)
                           \left(\frac{d\Phi}{d\bar r} 
                           - L(\bar{r})\right)~. \label{e:mod_TOV}
\end{eqnarray} 
$L(\bar r)$ denotes here the Lorentz force contribution, which is
noted $dM/dr$ in~\textcite{bonazzola-93} (see this reference for more
details). Note that these equations~(\ref{e:mod_TOV}) are not derived
from any first principle, but only motivated by Eq.(\ref{eq:bowers}),
in which we have replaced the diverging term by some phenomenological
term, supposed to better take into account the magnetic pressure and
the Lorentz force acting on the fluid.
\begin{figure*}
      \includegraphics[width=\columnwidth]{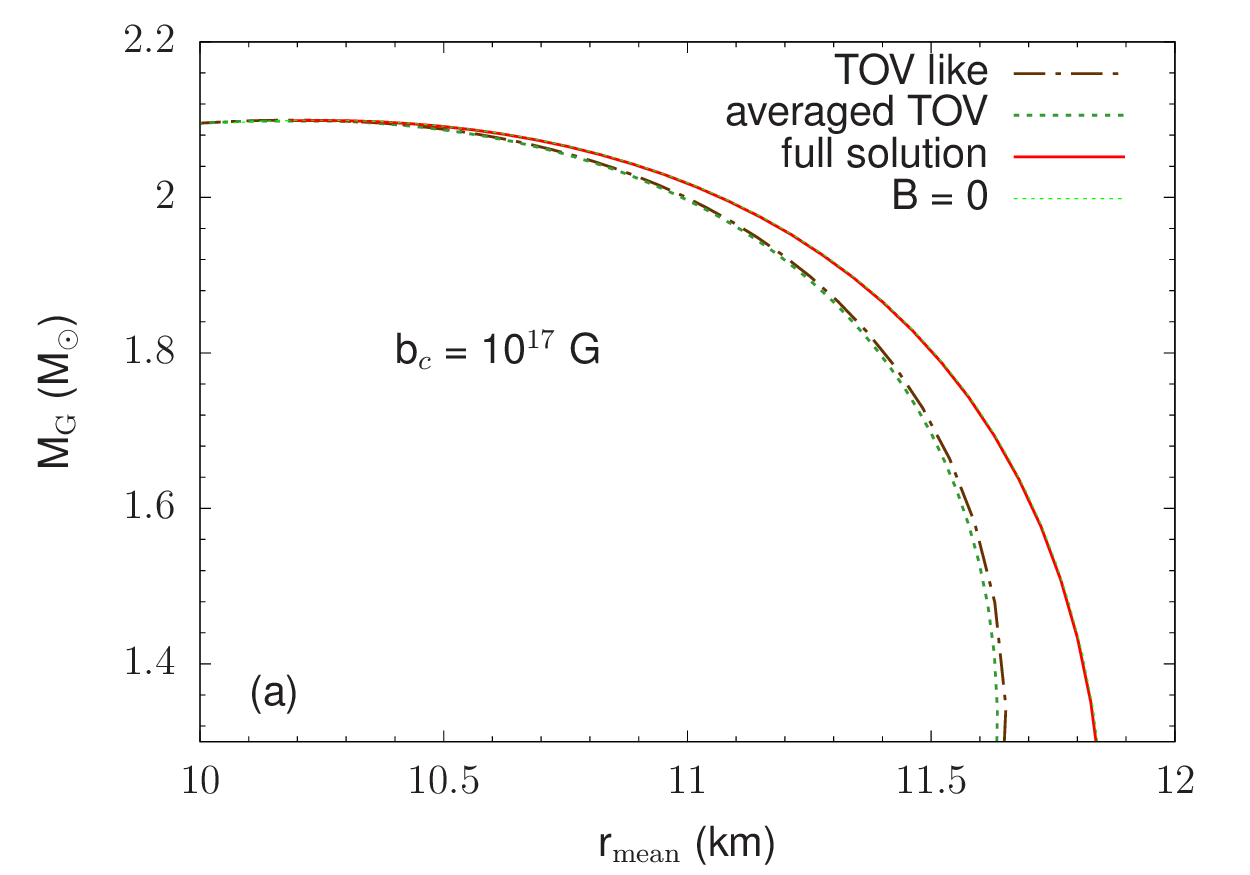} \hfill
      \includegraphics[width=\columnwidth]{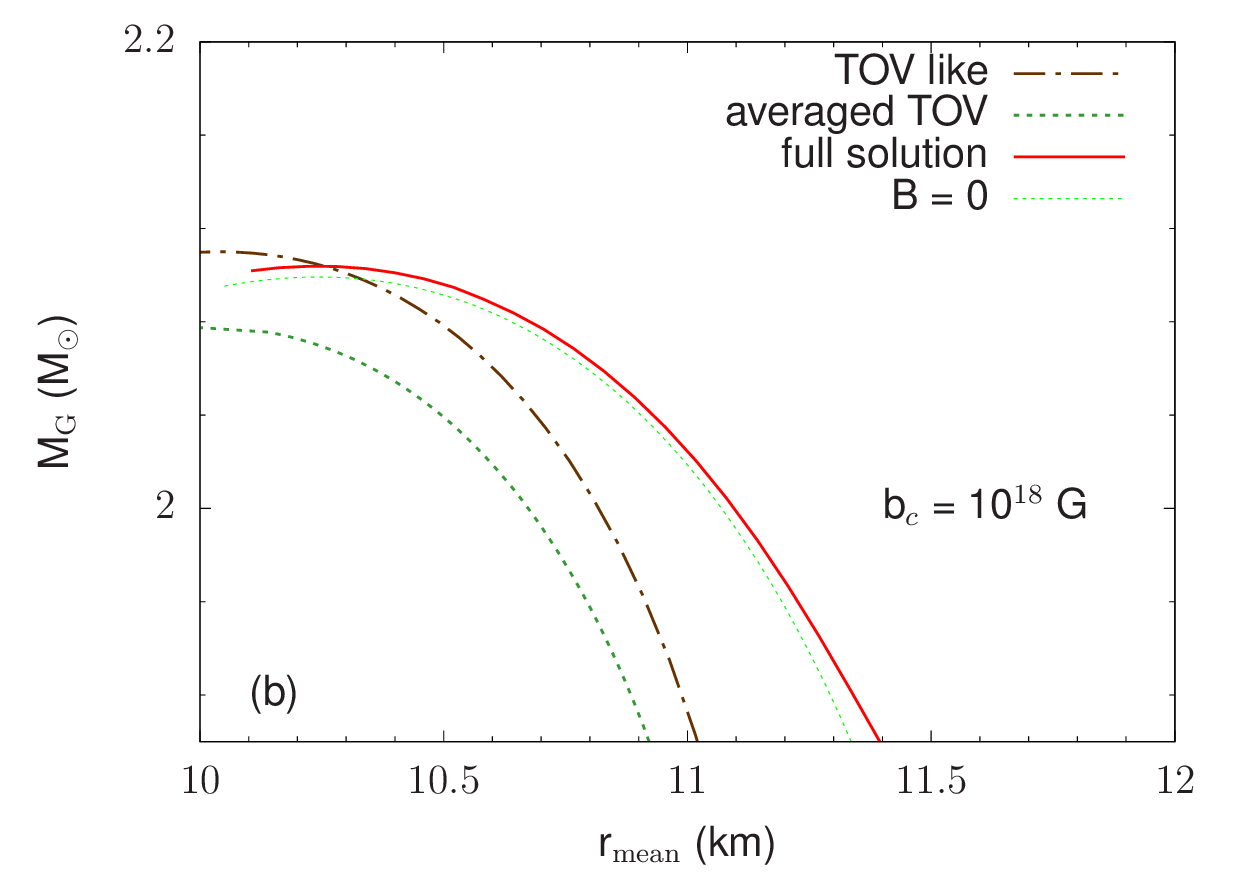}
      \caption{(color online) Gravitational mass vs. mean radius
        stellar sequences for the TOV-like solution, TOV solution
        using a directionally-averaged energy momentum tensor (see
        text) and the exact 
        (axisymmetric) calculation for a central magnetic field of
        $b_c = 10^{17}$G (left) and $b_c = 10^{18}$G (right) employing
        the SLy230a EoS. Note that, in both figures, the thin-dotted
        lines ($B=0$) correspond to solutions of the TOV system with
        no magnetic field at all. 
        \label{fig:tovmr}}
\end{figure*}

Similar to the magnetic field norm $b$, we found from the full
numerical calculations the following parametric form
\begin{equation}
L(\bar r) = 10^{-41}\times b_c^2 \left(-3.8\,x + 8.1\, x^3- 1.6\, x^5
  - 2.3\, x^7\right)~,
\end{equation}
where $x = \bar r/r_{\mathit{mean}}$ and the central magnetic field,
$b_c$, is given in units of G. For the magnetic field $b$ in
Eqs.~(\ref{e:mod_TOV}), the profile~(\ref{eq:bprofile}) is applied.

\begin{figure*}
  \begin{center}
      \includegraphics[width=\columnwidth]{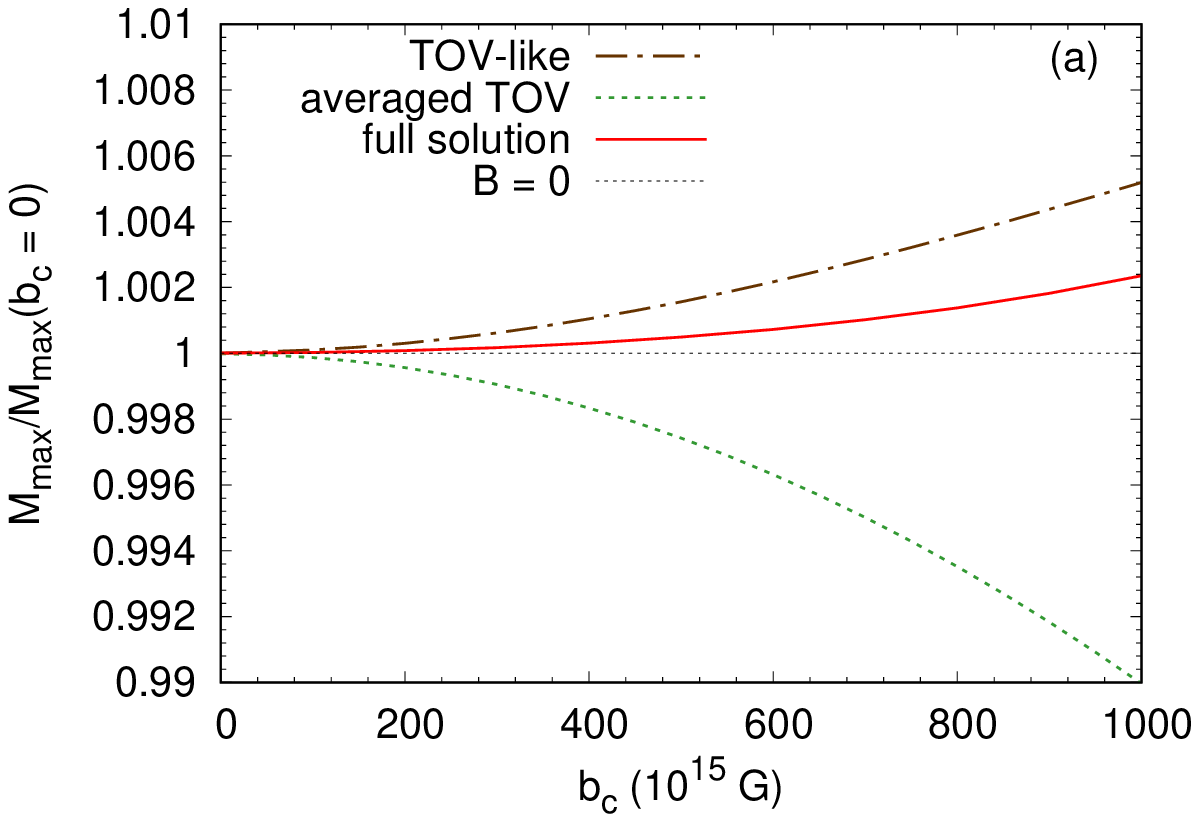}\hfill
      \includegraphics[width=\columnwidth]{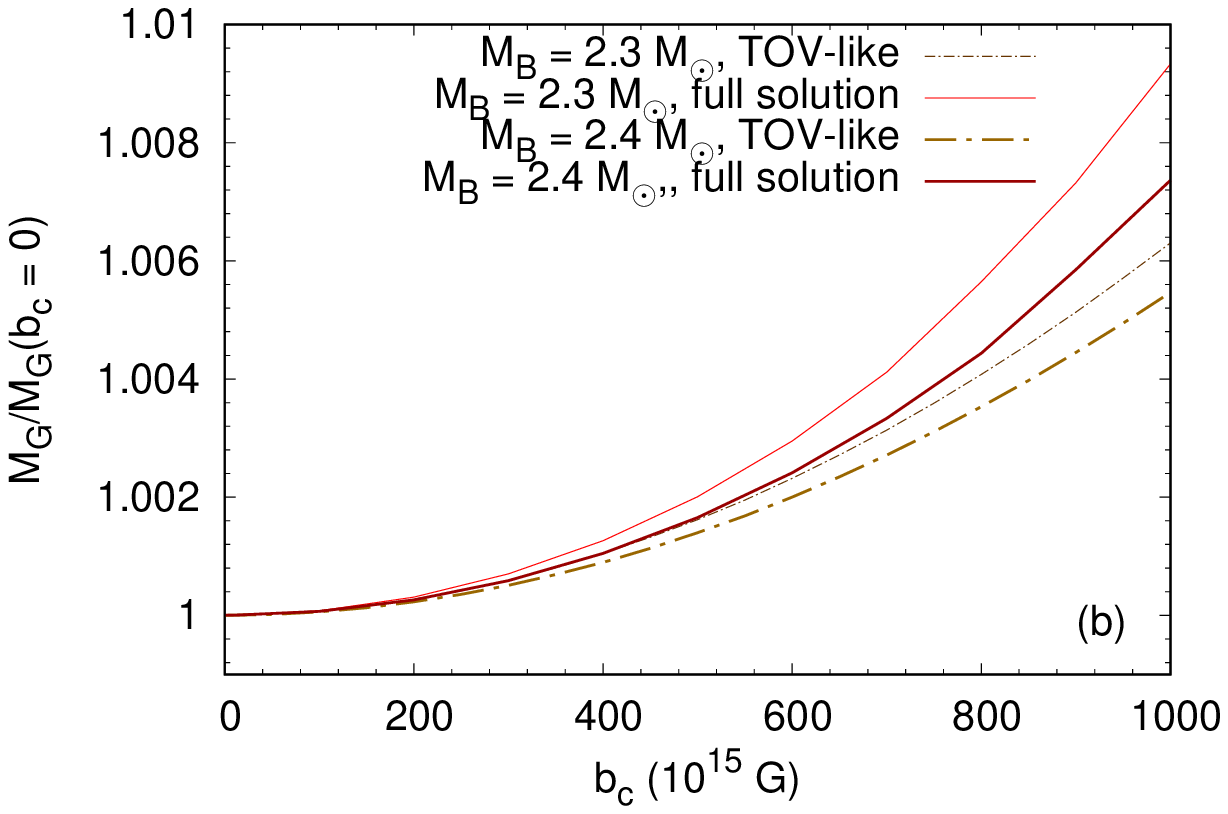}
    \caption{(color online) Left: Comparing the values of the maximum
      gravitational mass as function of the central magnetic field
      between our TOV-like approach, the averaged TOV and the full solution. Right:
      Gravitational mass as function of the central magnetic field for two
      different values of the baryonic mass.
      \label{fig:tovresults}}
  \end{center}
\end{figure*}

In order to get an idea of the quality of this ``TOV-like'' approach,
we show in Figs.~(\ref{fig:tovmr},\ref{fig:tovresults}) a comparison
between stellar sequences (described in mass vs. radius diagrams)
obtained with the TOV-like approach in spherical symmetry and the full
numerical solution in axial symmetry. In Fig.~(\ref{fig:tovmr}) the
gravitational mass vs. the mean radius is displayed for a central
magnetic field of $b_c = 10^{17}$~G (l.h.s.)  and $b_c = 10^{18}$~G
(r.h.s). As expected, deviations become larger at smaller masses since
the ratio of magnetic to matter pressure increases and thus the stars
are more strongly deformed and the relation between the mean radius
and the radius of a spherically symmetric configuration is no longer
obvious. The same is true if the central magnetic field is
increased. For further comparisons, we show in addition the solutions
obtained with TOV applying the ``averaged'' shear stress
tensor~\cite{bednarek-03}, mentioned in Sec.~\ref{s:intro}, and those
obtained without any magnetic field ($B=0$), too. We display here
results from the former approach since it has been used several times
in the literature~\cite{bednarek-03,dexheimer-12,menezes-16} for
studying properties of magnetized stars. Let us emphasize, however,
that the procedure to obtain the ``averaged'' stress-energy tensor is
mathematically ill-defined, as it is not possible to directionally
``average'' elements of a tensor. Moreover, looking at
Fig.~\ref{fig:tovmr} results with this approach for the mass-radius
relation most strongly differ from the full solution. Surprisingly,
the usual TOV solution with no magnetic field ($B=0$), on the
contrary, very well reproduces the full solution. Thus, although the
star becomes strongly deformed, the mean radius is only marginally
influenced by the magnetic field.

Masses should be less sensitive to the deformation. As can be seen in
Fig.~(\ref{fig:tovresults}), indeed the TOV-like solution for
the maximum gravitational mass as well as the gravitational mass for fixed
baryon mass as function of the central magnetic field show the correct
qualitative behavior. Both increase with $b_c$ and the TOV-like
approach overestimates the masses up to a factor two in the correction
(with respect to the non-magnetized case) at central fields of
$b_c = 10^{18}$G. This difference shows the difficulty of
``spherically-symmetric'' approaches to model magnetized neutron
stars. The limits of the TOV-like approach can be clearly seen in the
determination of effects due to the magnetic field. Finally, note that the
``averaged'' TOV approach gives a wrong qualitative behavior for the
dependence of the maximum mass on the central magnetic field value.

Thus, although our investigations can serve as a guideline and
reproduce at least for gravitational mass as function of magnetic
field the correct qualitative tendency, it should be stressed that it
is strongly recommended to use some consistent axisymmetric or
three-dimensional approach (\textit{e.g.} employing publicly available
software), to determine properties of magnetized neutron stars or to
draw any quantitative conclusion.
\section{Conclusions}\label{s:conc}

Many attempts can be found in the literature trying to study strongly
magnetized neutron stars and to include magnetic field effects on the
matter properties. As mentioned in the introduction, most of these
investigations suffer from different assumptions and approximations
motivated by the complexity of the full system of equations. First, in
order to avoid solving Maxwell's equations in addition to equilibrium
and Einstein equations, often an \textit{ad hoc} profile for the
magnetic field is assumed, which has no physical motivation. Second,
spherical symmetry is assumed for modelling the star.

In this work, we tackle the first point: we proposed a ``universal''
parameterization of the magnetic field profile (Eq.~\ref{eq:bprofile})
as a function of dimensionless stellar radius, obtained from a full
numerical calculation of the magnetic field distribution. We tested
this profile against several realistic hadronic EoSs, based on
completely different approaches, and with different magnetic
field strengths in order to confirm its universality. For the case of
quark matter EoSs, preliminary investigations showed that although MIT
bag models conform to the universality, other quark matter EoSs may
not necessarily do so. The profile is intended to serve as a tool for
nuclear physicists for practical purposes, namely to obtain an
estimate of the maximum field strength as a function of radial depth,
within the error bars observed in our study, \textit{e.g.} in
Fig.~(\ref{f:uprof}), in order to deduce the composition and related
properties.

We applied the proposed magnetic field profile in a modified TOV-like
system of equations, that include the contribution of magnetic field
to the energy density and pressure, and account for the anisotropy by
introducing a Lorentz force term. Compared with full numerical
structure calculations, we find that qualitatively the correct
tendency is reproduced and quantitatively the agreement is acceptable
for large masses and small magnetic fields ($b_c \lesssim
10^{17}$~G). However, we find that the standard TOV system with no
magnetic field reproduces much better mass-radius relations, even for
strong magnetic fields, than any modified TOV system, with poorly
defined magnetic corrections. This is mostly due to the fact that the
mean radius is only marginally changed by the magnetic field. We thus
think that future studies should employ the profile proposed here to
conclude about the importance of magnetic field effects on matter
properties, and use TOV system at $B=0$ for calculating mass-radius
diagrams. For any other property of magnetized stars, we can only
recommend the use of a full axisymmetric numerical solution for
modelling magnetized neutron stars.


\begin{acknowledgments}
  This work was financially supported by the action ``Gravitation et
  physique fondamentale'' of Paris Observatory. DC acknowledges
  financial support from CNRS and technical support from LPC Caen and
  the Paris Observatory in Meudon where part of this study was
  completed.
\end{acknowledgments}
\bibliography{biblio}

\end{document}